\documentclass{llncs}

\usepackage{amsmath}
\usepackage{amsfonts}
\usepackage{amssymb}

\usepackage{url}
\usepackage{hyperref}
\hypersetup{
  colorlinks=true,
  linkcolor=blue,
  citecolor=green
}
\usepackage[nameinlink]{cleveref}
\crefname{table}{table}{tables}
\Crefname{table}{Table}{Tables}
\crefname{figure}{figure}{figures}
\Crefname{figure}{Figure}{Figures}
\crefname{section}{section}{sections}
\Crefname{section}{Section}{Sections}
\crefname{claim}{claim}{claims}
\Crefname{claim}{Claim}{Claims}

\usepackage[usenames,dvipsnames]{xcolor}
\usepackage{graphicx}
\usepackage{caption}

\usepackage{ytableau}

\usepackage{enumerate}

\spnewtheorem*{theorem*}{Theorem}{\bfseries}{\rmfamily}
\spnewtheorem{claim}{Claim}{\itshape}{}

\newcommand{\Ext}{\mathsf{Ext}}

\newcommand{\cH}{\mathcal{H}}
\newcommand{\Var}{\mathrm{Var}}

\newcommand{\SD}{\mathrm{SD}}

\newcommand{\cX}{\mathcal{X}}

\newcommand{\bigOm}[1]{\Omega\left(#1\right)}
\newcommand{\bigO}[1]{O\left(#1\right)}

\DeclareMathOperator*{\E}{\mathbb{ E }}
\DeclareMathOperator*{\Hmin}{\mathbf{ H }_{\infty}}

\title{Lower bounds on $q$-wise independence tails and applications to min-entropy condensers}

\author{Maciej Sk\'{o}rski 
\thanks{This work was partly supported by the WELCOME/2010-4/2 grant founded within the framework of the EU Innovative Economy Operational Programme.} }
\institute{
\email{maciej.skorski@mimuw.edu.pl} \\ Cryptology and Data Security Group, University of Warsaw}
\begin{document}

\maketitle

\begin{abstract}
We present novel and sharp lower bounds for higher load moments in the classical problem of mapping $M$ balls into $N$ bins by $q$-universal hashing, specialized to the case when $M=N$. As a corollary we prove a tight counterpart for the result about min-entropy condensers due to Dodis, Pietrzak and Wichs (CRYPTO'14), which has found important applications in key derivation. It states that condensing $k$ bits of min-entropy into a $k$-bit string $\epsilon$-close to almost full min-entropy (precisely $ k-\log\log(1/\epsilon)$ bits of entropy) can be achieved by the use of $q$-independent hashing with $q= \log(1/\epsilon)$. We prove that when given a source of min-entropy $k$ and aiming at entropy loss $\ell = \log\log (1/\epsilon) - 3$, the independence level $q=(1-o(1))\log(1/\epsilon)$ is necessary (for small values of $\epsilon$), which almost matches the positive result. Besides these asymptotic bounds, we provide clear hard bounds in terms of Bell numbers and some numerical examples. Our technique is based on an explicit representation of the load moments in terms of Stirling numbers, some asymptotic estimates on Stirling numbers and a tricky application of the Paley-Zygmund inequality. 

\end{abstract}

\keywords{min-entropy condensers, key derivation, balls and bins hashing, anti-concentration inequalities}

\newpage

\section{Introduction}

\subsection{Universal hashing and key derivation}

Random variables $\xi_1,\ldots,\xi_N$ are called $q$-wise independent if every $q$ of them are fully independent. $p$ wise independence find important applications in cryptography, for example in constructing pseudorandom generators \cite{HILL88}, oblivious transfer protocols \cite{Bellare1994}, or key derivations \cite{DodisPW14}. In this work we focus on the last area, where most recent results offer a huge improvement by replacing general purpose randomness extractors by randomness condensers based on independent hashing.

\subsection{Better key derivation by independent hashing}

A min-entropy condenser is a primitive which transforms a distribution with some entropy (of a possibly small rate) into a distribution of almost full entropy.
Dodis et al. prove the following theorem
\begin{theorem}[Parameters for $q$-universal condensers, \cite{DodisPW14}]\label{thm:condensors_dodis}
Any $q$-universal family from $n$ to $m$ bits is a $(k,\ell,\epsilon)$-condenser with
$k=m$, $q=\log(1/\epsilon)$, $\ell=\log q$.
\end{theorem}
Informally it states that by $q$-independent hashing we condense a possibly long string of min-entropy $k$ into a $k$-bit string $2^{-q}$-close (in the statistical distance) to a $k$-bit string of entropy $k-\log q$, which is almost full. This technical result is a key ingredient of their important work on key derivation. The second is the important observation that for a wide class of so called unpredictability applications one can use a weak key with only high entropy, achieving roughly the same quality as with a key close to uniform. Combining these two facts they are able to reduce the entropy loss from $2\log(1/\epsilon)$, offered by general purpose extractors (necessary by the RT-bound \cite{Radhakrishnan2000}), to roughly $\ell = \log\log(1/\epsilon)$, offered by $q$-independent condensers where $q=\log(1/\epsilon)$, when the required security strength is $\epsilon$. The higher independence level $q$, the better security guarantees we get. However, too big $q$ affects the efficiency of computations (time complexity) and the sampling cost (the need of longer seeds) of the hashing family. Thus the following question is natural
\begin{quote}
\textbf{Q}: Suppose we use $q$-wise independent hashing to condense a source of min-entropy $k$ into an $m$-bit key close to have min-entropy almost $m$. What is the minimal value of  $q$?
\end{quote}
The following question is stated informally, depending on what we understand by ``close'' and ``almost full''.  Taking a positive result as a reference point, we set $k=m$ and allow the derived key to be $\epsilon$-close to an $k$-bit source of $k-\log\log(1/\epsilon)$ min-entropy. A key with this quality ensures total security roughly $\epsilon/\log(1/\epsilon)$ for any unpredictability application \cite{DodisPW14}.

 

\subsection{Our contribution}

\textsc{Summary.} We show that the parameters for min-entropy condensers stated in \Cref{thm:condensors_dodis}
 are essentially tight. Our approach is based on a novel anticoncentration inequality derived by a Paley-Zygmund trick, which involves higher moments of load in a balls-bins problem. We also use some bounds on Stirling numbers of  second kind to simplify expressions describing load moments, into more compact forms.
\vspace{2mm}\newline\textsc{Moments of independent boolean sums.}  We state the following useful fact, noticing that similar observations have been already exploited by some authors (see for instance \cite{berend2010improved} for a one-sided version of this inequality).
\begin{proposition}[Explicit moments of balls-bins loads]\label{thm:load_moments_explicit}
Let $S = \sum_{i=1}^{M} \xi_i$ be a sum of $q$-wise independent boolean random variables with mean $\E \xi_i = \frac{1}{N}$. Then we have the following identities
\begin{align}
 \E S^{q}  & = \sum_{j} S(q,j) N^{-j} \binom{M}{j} j! \label{eq:balls_bins_raw_moment} 
\end{align}
where $S(q,j)$ and 
denote 
Stirling numbers of the second kind. 
\end{proposition}
\begin{remark}[A balls-bins statement]
Think of $M$ balls, $N$ bins, and $\xi_i$ indicating whether the $i$-th ball is mapped into a chosen bin. Then $S$ is precisely the load of the bin.
\end{remark}
\vspace{2mm}\textsc{Anti-concentration bounds for boolean sums.} Our main tool is the following novel anti-concentration inequality for $q$-wise independent hashing
\begin{lemma}[Anti-concentration of balls-bins loads when $M=N$]\label{lemma:load_anticoncentration}
Let $S$ be as in \Cref{thm:load_moments_explicit} and $q\geqslant 4$ be an even number. Then we have the following inequality
\begin{align}\label{eq:anti_concentration}
\Pr\left[ S \geqslant \frac{\left( B_{\frac{q}{2}} \right)^{\frac{2}{q}}}{2} \right] \geqslant \left(1-(2M)^{-1}q^2\right) \cdot \frac{\left(  B_{\frac{q}{2}} \right)^2 }{ 2 B_q } 
\end{align}
where $B_q$ are Bell numbers.
\end{lemma}
\textsc{Impossibilities for min-entropy $q$-wise independent condensers.} Based on the previous lemma and deriving some estimates on $B_q$ we finally obtain
\begin{theorem}[Impossibilities for $q$-universal condensers]\label{thm:condensers_lower_bounds}
No $q$-universal family from $n$ to $m$ bits can be a $(k,\ell,\epsilon)$-condenser where
\begin{align*}
k &=m \\
  \ell &= \log q-\log\log q -\log(2\mathrm{e}) + O\left(\frac{\log\log q}{\log  q} \right) \\
   \epsilon & = 2^{-q \left(1+ O\left(\frac{\log \log q}{ \log q} \right) \right)},
\end{align*} 
provided that $k > 2\log q$. Equivalently, it is not a $(k,\ell,\epsilon)$-condenser when
\begin{align*}
k &=m \\
  \ell &= \log \log(1/\epsilon) -\log\log \log(1/\epsilon) -\log(2\mathrm{e}) + O\left(\frac{\log\log \log(1/\epsilon)}{\log  \log (1/\epsilon)} \right) \\
   q & = \log(1/\epsilon)-\omega\left( \frac{\log(1/\epsilon)\cdot \log\log \log(1/\epsilon)}{\log  \log (1/\epsilon)}  \right),
\end{align*} 
provided that $m > 2\log\log(1/\epsilon)$. These facts are even true for all flat $k$-sources. 
\end{theorem}
Note that the additional assumption $k > 2\log q$ with $q=\log(1/\epsilon)$ is trivially satisfied for all practical applications. For recommended security $\epsilon = 2^{-80}$ it becomes $k \geqslant 13$. Since $k=m$, it satisfied even for condensing into only $m=13$ bits! We also stress that the result in \Cref{thm:condensers_lower_bounds} is asymptotically tight, but better hard bounds can obtained by using \Cref{lemma:load_anticoncentration} directly. For example, setting $q = 64$ doesn't yield a condenser with loss $\ell = 2.6$ and quality $\epsilon = 2^{-43}$, whereas the positive result yield a condenser with $\ell = 6$ and $\epsilon = 2^{-64}$.

\section{Preliminaries}

\textsc{Entropy and Statistical Closeness.} We say that $X$ has $k$ bits of entropy if $\Pr[X=x]\leqslant 2^{-k}$ for every $x$ in the range of $X$. Alternatively, we call $X$ a $k$-source. The statistical distance of $X_1,X_2$ is defined by $\mathrm{SD}\left(X_1;X_2\right) = \frac{1}{2}\sum_{x}\left| \Pr[X_1=x]-\Pr[X_2=x]\right|$; we also say that $X_1$ and $X_2$ are $\epsilon$-close. A $n$ bit key $X$ is $\epsilon$-secure if it is $\epsilon$-close to the uniform distribution over $n$-bit strings. In practice we think of $\epsilon = 2^{-80}$ as small enough to offer good security (indistinguishably).
\vspace{2mm}\newline\textsc{Independent Hashing.} A family $\{h_s\}_{s\in \{0,1\}^d}$ of functions from $n$ to $m$ bits is called $q$-wide independent hash family (or simply $q$-universal) if for any choice of distinct $n$-bit strings $x_1,\ldots,x_q$ and a randomly chosen $s$ the random variables $h_s(x_1),\ldots,h_s(x_q)$ are independent. This concept is due to Carter and Wegman \cite{Carter1977}.
\vspace{2mm}\newline\textsc{Combinatorial Numbers.} The Bell number $B_q$ counts the total number of partitions of a $q$-element set. The Stirling number of second kind $S(q,j)$ counts the number of partitions of a $q$-element set into precisely $j$ blocks.
\vspace{2mm}\newline\textsc{Sources and Condensers.} A distribution $X$ is called $k$-source if it has min-entropy at least $k$. The following definition formalizes the notion of min-entropy condensers, whose purpose is to increase the entropy rate (density)
\begin{definition}[Min-entropy condensers]
A function $\mathsf{Cond}:\{0,1\}^n\times\{0,1\}^d\rightarrow \{0,1\}^m$ is a $(k,\ell,\epsilon)$-condenser with a $d$-bit seed if for any $k$-source $X$ and a randomly chosen $s\in\{0,1\}^d$ the distribution of $\mathsf{Cond}(X,s)$ is $\epsilon$-close to some distribution of $m-\ell$ bits of min-entropy.
\end{definition}

\section{Proofs}

\subsection{Proof of \Cref{lemma:load_anticoncentration}}

Recall the standard Paley-Zygmund inequality
\begin{align}\label{eq:PaleyZygmund}
 \Pr[ Y \geqslant \theta \E Y ] \geqslant (1-\theta)^2\cdot \frac{(\E Y)^2}{\E Y^2},\quad 0<\theta < 1
\end{align}
valid for arbitrary non-negative $Y$. Setting $Y = \E S^{\frac{q}{2}}$ we obtain
\begin{align}\label{eq:anti_concentration_1}
\Pr\left[ S \geqslant \theta^{\frac{2}{q}}  \left\| S\right\|_{\frac{q}{2}} \right] \geqslant (1-\theta)^2\cdot \frac{\left(  \left\| S\right\|_{\frac{q}{2}}\right)^{q} }{  \left( \left\| S\right\|_{{q}} \right)^q} 
\end{align}
Note that for the special case $M=N$ by \Cref{thm:load_moments_explicit} we obtain 
\begin{align*}
\prod_{i=1}^{q}\left(1-\frac{i-1}{M}\right)\cdot \sum_{j} S(q,j) \leqslant \E S^{q} \leqslant \sum_{j} S(q,j) 
\end{align*}
Since we have $\prod_{i=1}^{q}(1-a_i) \geqslant 1-\sum_{i=1}^{q} a_i$ (an elementary inequality provable by induction),  \Cref{eq:anti_concentration_1} specializes to
\begin{align}\label{eq:anti_concentration_2}
\Pr\left[ S \geqslant \theta^{\frac{2}{q}} \left( B_{\frac{q}{2}} \right)^{\frac{2}{q}} \right] \geqslant (1-\theta)^2 \left(1-(2 M)^{-1}q^2\right)\cdot \frac{\left(  B_{\frac{q}{2}} \right)^2 }{ B_q } 
\end{align}
Setting $\theta = \frac{1}{q}$ we obtain $\theta^{\frac{2}{q}} \geqslant \frac{1}{2}$ and $(1-\theta)^2 \geqslant \frac{1}{2}$.\qed

\begin{lemma}[Maximum of Stirling numbers of second kind, \cite{Rennie1969116}]\label{lemma:Stirling2max_bounds}
We have 
\begin{align}
\frac{\max_{j} \ln S(q,j)}{q} = \ln q - \ln\ln q -1 + O\left(\frac{\ln\ln q}{\ln q}\right)
\end{align}
as $q\rightarrow +\infty$.
\end{lemma}
By noticing that $\max_{j} \ln S(q,j) < B_j q\cdot \max_{j} \ln S(q,j)$ we obtain the following corollary about the growth rate of Bell numbers.
\begin{corollary}[Bounds on Bell numbers]\label{lemma:Bell_bounds}
We have
\begin{align}
\frac{\ln B_q}{q} = \ln q - \ln\ln q -1 + O\left(\frac{\ln\ln q}{\ln q}\right)
\end{align}
as $q\rightarrow +\infty$.
\end{corollary}

\subsection{Proof of \Cref{thm:condensers_lower_bounds}}

\begin{proof}
Since $1-M^{-1}q^2 \geqslant \frac{1}{2}$, we need to ensure that 
\begin{align*}
  \frac{\left(B_{\frac{q}{2}}\right)^2}{B_q} \geqslant 4\epsilon
\end{align*}
which is equivalent to
\begin{align}\label{eq:anticoncentration_concrete_1}
  \frac{\ln B_{\frac{q}{2}}}{\frac{q}{2}} - \frac{\ln B_q}{q} \geqslant \frac{\ln (4\epsilon) }{q}.
\end{align}
By \Cref{lemma:Bell_bounds}, this inequality means
\begin{align*}
 \ln 2 \geqslant  \frac{\ln (4\epsilon) }{q} + O \left(\frac{\ln\ln q}{\ln q}\right) ,
\end{align*} 
which expressed in logarithms at base $2$ is equivalent to
\begin{align*}
 \frac{\log (1/\epsilon) }{q} \geqslant 1 + O \left(\frac{\log\log q}{\log q}\right).
\end{align*}
By taking the inverses and using the Taylor series expansion $\frac{1}{1+x} \approx 1-x$ for $x\approx 0$ we can rewrite it as
\begin{align}\label{eq:anticoncentration_concrete_2}
 \frac{q}{\log (1/\epsilon)} \leqslant 1 - O \left(\frac{\log\log q}{\log q}\right).
\end{align}
Thus, it suffices to find possibly good $q$ such that
\begin{align}\label{eq:anticoncentration_concrete_2}
 \frac{q}{\log (1/\epsilon)} + c \cdot \left(\frac{\log\log q}{\log q}\right) \leqslant 1
\end{align}
where $c$ a positive constant (comparable up to a small constant factor to the absolute value of the constant hidden in \Cref{lemma:Bell_bounds}). Take $q=(1-\gamma)\log(1/\epsilon) $ where the exact value of $\gamma$ is to be determined. Since
for $q\geqslant 8$ the function $q\rightarrow \frac{\log\log q}{\log q}$ is decreasing, we see that it suffices to satisfy
\begin{align*}
 -\gamma + c\cdot \frac{\log\log \log(1/\epsilon)}{\log\log (1/\epsilon)} \leqslant 0
\end{align*}
and therefore we put 
\begin{align*}
 \gamma =c\cdot \frac{\log\log \log(1/\epsilon)}{\log(1/\epsilon)\cdot\log\log (1/\epsilon)},
\end{align*}
which finishes the proof if we set $M=N=2^{k}$. \qed
\end{proof}

\section{Conclusions}
It would be interesting to extend the results to settings when $M > N$. We leave this as an open problem for further research. We are also going to extend our methods to cover the case of \emph{almost independent} hash functions which are most suitable in practical implementations because of much shorter seeds \cite{DodisPietrzakWichs2013}.

\bibliographystyle{amsalpha}
\bibliography{./citations}

\end{document}